\documentclass{aa}
\usepackage{txfonts}
\usepackage{natbib}
\usepackage{graphicx}

\bibpunct[, ]{(}{)}{;}{a}{}{,}

\def\llm{{\sc LLmodels}}
\def\atl{{\sc ATLAS9}}

\def\logg{\log g}
\def\teff{T_{\rm eff}}

\def\kms{km\,s$^{-1}$}

\def\paper1{Paper~I}
\def\hbeta{H$\beta$}
\def\hgamma{H$\gamma$}

\begin{document}

\title{Model atmospheres of magnetic chemically peculiar stars}
\subtitle{A remarkable strong-field Bp SiCrFe star HD\,137509}

\author{D. Shulyak\inst{1} \and O. Kochukhov\inst{2} \and S. Khan\inst{3,4} }
\offprints{D. Shulyak, \\
\email{denis@jan.astro.univie.ac.at}}
\institute{Institute of Astronomy, Vienna University, Turkenschanzstrasse 17, 1180 Vienna, Austria \and
Department of Astronomy and Space Physics, Uppsala University, Box 515, 751 20, Uppsala, Sweden \and
Physics and Astronomy Department, University of Western Ontario, London, ON, N6A 3K7, Canada \and
Institute for Computational Astrophysics, Saint Mary's University, 923
Robie Street, Halifax, B3H 3C3, Nova Scotia, Canada
}

\date{Received / Accepted}

\abstract
{In the last few years we have developed stellar model atmospheres which included
effects of anomalous abundances and strong magnetic field.
In particular, the full treatment of anomalous Zeeman splitting and polarized radiative
transfer were introduced in the model atmosphere calculations for the first time.
The influence of the magnetic field
on the model atmosphere structure and various observables
were investigated for stars of different fundamental parameters and metallicities.
However, these studies were purely theoretical and did not attempt to model real objects.}
{In this investigation we present results of modelling the atmosphere of one of the most
extreme magnetic chemically peculiar stars, HD\,137509. This Bp SiCrFe star has the mean surface
magnetic field modulus of about $29$\,kG.
Such a strong field, as well as clearly observed abundance peculiarities, make this star
an interesting target for application of our newly developed model atmosphere code.}
{We use the recent version of the line-by-line opacity sampling stellar model atmosphere code \llm,
which incorporates the full treatment of Zeeman splitting of spectral lines, detailed
polarized radiative transfer and arbitrary abundances. We compare model predictions with photometric
and spectroscopic observations of HD\,137509, aiming to reach a self-consistency between the
abundance pattern derived from high-resolution spectra and abundances used for model atmosphere calculation.
}
{Based on magnetic model atmospheres, we redetermined abundances and fundamental parameters of
HD\,137509 using spectroscopic
and photometric observations. This allowed us to obtain a better
agreement between observed and theoretical parameters compared to non-magnetic models with
individual or scaled-solar abundances.}
{We confirm that the magnetic field effects lead to noticeable changes in the model atmosphere structure
and should be taken into account in the stellar parameter determination and abundance analysis.}


\maketitle

\section{Introduction}
\label{intro}
The atmospheric structure of magnetic chemically peculiar (CP) stars deviates from that of normal
stars with similar fundamental parameters due to unusual chemistry, abundance
inhomogeneities and the presence of strong magnetic field. These effects are not considered
in the standard model atmosphere calculations, possibly leading to errors in
the stellar parameter determination and abundance analysis. To circumvent this long-standing
problem of stellar astrophysics, we have developed a new line-by-line opacity sampling model
atmosphere code \llm\ \citep{llm}. Using this tool
in the series of recent papers, we investigated in detail the effects of
anomalous Zeeman splitting \citep{zeeman_paper1}, polarized radiative transfer \citep{zeeman_paper2}
and inclination of the magnetic field vector \citep{zeeman_paper3} on the model structure,
energy distribution, hydrogen line profiles, photometric colors and the magnitude
of bolometric corrections for a grid of model atmospheres with different effective
temperatures and metallicities. For the first time we were able to obtain
new results applying direct and self-consistent modeling of all effects mentioned above
and to answer the question
how does the magnetic field act at different temperatures and what one could expect
if the magnetic field is ignored in calculations of model atmosphere of magnetic CP stars.
It was shown that the strength of the magnetic field
is the key characteristic controlling the magnitude of the magnetic field effects and
the polarized radiative transfer should be taken into account.
In contrast, the orientation of the magnetic field vector does not have much influence
on any of the observed stellar characteristics and, thus, can be safety ignored in the analysis
routines.

So far our models with the magnetic field effects included have been developed and applied
only in the context of purely theoretical studies. Here we make the first attempt
to model the atmosphere of a real star.
In this work we use the \llm\, stellar model atmosphere code
to investigate the atmospheric structure of the star, taking into account
individual chemical composition, anomalous Zeeman splitting
and polarized radiative transfer.

HD\,137509 (HIP\,76011, NN~Aps) is a B9p SiCrFe chemically peculiar star with a strong reversing
longitudinal field and variable lines of \ion{Si}{ii} and iron-peak elements
\citep{mathys91,mathys_lanz}. \citet{paper1} (hereafter \paper1) has detected resolved Zeeman split
lines in the spectrum of HD\,137509, showing that this star is characterized by a non-dipolar
magnetic field geometry with a mean surface field strength of  about $29$\,kG. This is the
second-largest magnetic field ever found in a CP star (the first place is occupied by the well-known
Babcock's star, \citet{preston}). The atmospheric parameters, $\teff=12\,750\pm500\,K$ and $\log g=3.8\pm0.1$, were
derived in \paper1\ using theoretical fit to the observed \hbeta\, and \hgamma\, line profiles based on the
\atl\, \citep{kurucz13} model with enhanced metallicity, $[M/H]=+1.0$.
The appearance of such a strong magnetic field and the presence of outstanding abundance
anomalies inferred in \paper1\ allow us to use HD\,137509 as a test ground for the
application of the new generation magnetic model atmospheres.

In the next section we briefly describe the techniques employed to construct magnetic model
atmospheres. Results of the calculations for HD\,137509 are presented in Sect.~\ref{results}.
Main conclusions of our study are summarized in Sect.~\ref{concl}.

\section{Model atmosphere calculations}
\label{models}

To calculate magnetic model atmospheres we used the current version of the \llm\, code,
originally developed by \citet{llm}. The \llm\, is a LTE, 1-D, plane-parallel, hydrostatic stellar
model atmosphere code for early and intermediate-type stars. The direct line-by-line calculation
of the bound-bound opacities implemented in this code
allows one to account for arbitrary individual and stratified stellar
abundance patterns and include various effects caused by the magnetic field.
The VALD database \citep{vald1,vald2} was used as a source of atomic line parameters.
The extracted lines were subjected to a preselection procedure inside \llm, with the
default criterion $\alpha_{\rm line}/\alpha_{\rm cont}>1\%$ in at least one atmospheric layer
allowing to select only those lines that significantly contribute to the
opacity (here $\alpha_{\rm line}$ and $\alpha_{\rm cont}$ are the line center and continuum opacity
coefficients, respectively). The numerical experiments showed that this criterion
is sufficient for accurate representation of the energy distribution and T-P structure of the models. Decreasing
this value leads to significant increase of the number of selected lines but without producing noticeable
changes in overall blanketing \citep[see][]{llm}. The preselection
procedure was performed twice per model atmosphere calculation: at the first iteration
and at the iteration when the temperature correction at each model
atmosphere layer is less than few tens of K. The remaining model iterations are
performed with the later line list, thus ensuring a consistency of the preselected lines and the
model structure obtained. The spectrum synthesis for the magnetic model atmosphere calculations was
carried out in the range between
100 and 40\,000\,\AA, with a constant wavelength step of 0.1\,\AA.

Generally, due to the variation of magnetic field vector across the visible stellar surface, one has
to compute a number of local model atmospheres for each appropriately chosen surface grid element
using individual values of the strength and inclination of the magnetic field. Then, the total flux
coming from the star should be obtained by integration of the radiation field intensity
corresponding to individual surface zones. However, as it was shown by \citet{zeeman_paper3}, the
inclination of the magnetic field vector  does not significantly influence the structure of magnetic
models and the resulting energy  distribution, i.e. the anisotropy effects can be neglected in the
magnetic model atmosphere calculations. Consequently, we assume the magnetic field vector to be
perpendicular to the atmosphere normal and, according to the results of \paper1, adopt the field
strength of 29~kG.
The approach for  calculating magnetic models used here is equivalent to the one
described in Sect.~2 in \citet{zeeman_paper2}.

For all models presented in this study we assume a force-free configuration of the surface
magnetic field. This agrees with the results of multipolar modeling of magnetic topology
\citep{paper1} and implies that possible modification of the hydrostatic equilibrium by the
Lorentz force (see \citet{lorentz} and references therein) is absent in HD\,137509. Thus, all differences in the
pressure structure between magnetic and non-magnetic models that we obtain are caused only by additional opacity
in the Zeeman components.

The necessity to perform polarized radiative transfer calculations over a wide wavelength range
makes the models with magnetic field very computationally expensive. To reduce
computational costs, we used the following approach. Fixing fundamental parameters and abundance pattern
we calculated a non-magnetic model. After this model is converged, a model with
magnetic field is iterated using the temperature-pressure structure of the initial non-magnetic
model as a first approximation.

The abundances of seven chemical elements for HD\,137509 were derived in \paper1\
and are listed in Table~\ref{Tabn} second column). Due to the abnormal weakness of the He lines,
the value of the He abundance should be considered as an upper limit only.
One can note the strong overabundance of Si, Fe, Cr and Ti, whereas Ca and
Mg are underabundant by more than 1~dex relative to the solar chemical composition.
We assume solar abundances from \citet{asplund} for all other elements. Using available observations it
is not possible to derive abundances of elements other than those listed in Table~\ref{Tabn}
with good accuracy due to complex line shapes and strong blending by iron-peak elements. There are no
usable lines of intrinsically abundant elements such as C, N and Ne, while the abundance of O that could in
principle be determined from the IR triplet (7772--7775 \AA) is unreliable due to large NLTE effects and
instrumental artifact in the UVES spectrum of HD\,137509 in this region.  According to our recent investigation
\citep{indabn}, C, N, and O belong to the group of elements that produces only a small change in the model
structure of A and B stars even if their abundances are reduced by $1$~dex relative to the solar values.
Finally, to ensure the consistency between the model atmosphere and
the abundance pattern of the star,
we constructed stellar atmosphere models with these individual abundances including magnetic
field. Then we try to assess relative importance of this sophisticated approach.

\begin{table}
\caption{Element abundances of HD\,137509 compared to the
solar values \citep{asplund}. The second and the third columns give, respectively, abundances derived using
the scaled-solar abundance model atmosphere (\paper1) and using model with $\teff=13750$\,K, $\logg=4.2$
with magnetic field included (see Sec.~3.4).
Abundances are given in the logarithmic scale $\log(N_{\rm el}/N_{\rm total})$.}
\centering
\begin{tabular}{c|c|c|c}
\hline\hline
 Element & t12750g3.8 (\paper1) & t13750g42 & Sun\\
\hline
  He & $<-3.50$ & $<-3.50$ & $-1.10$\\
  Si & $ -3.73$ & $ -3.58$ & $-4.53$\\
  Fe & $ -3.19$ & $ -3.00$ & $-4.59$\\
  Cr & $ -4.20$ & $ -3.90$ & $-6.40$\\
  Ti & $ -4.54$ & $ -4.20$ & $-7.14$\\
  Ca & $ -7.93$ & $ -7.50$ & $-5.73$\\
  Mg & $ -5.71$ & $ -5.50$ & $-4.51$\\
\hline
\end{tabular}
\label{Tabn}
\end{table}

\section{Results}
\label{results}

\subsection{Model structure}

\begin{figure}
\includegraphics*[angle=-90,width=\hsize]{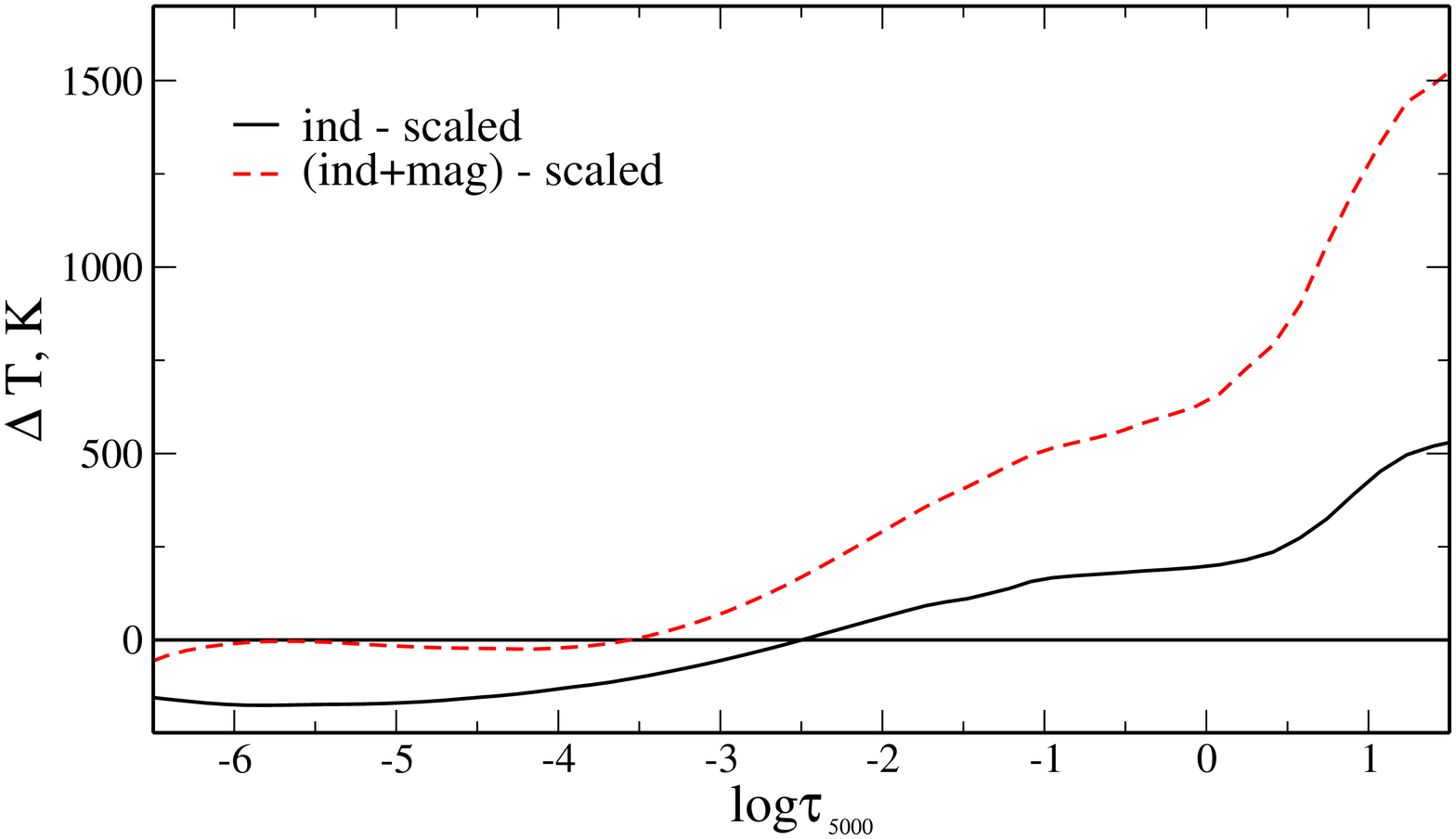}
\includegraphics*[angle=-90,width=\hsize]{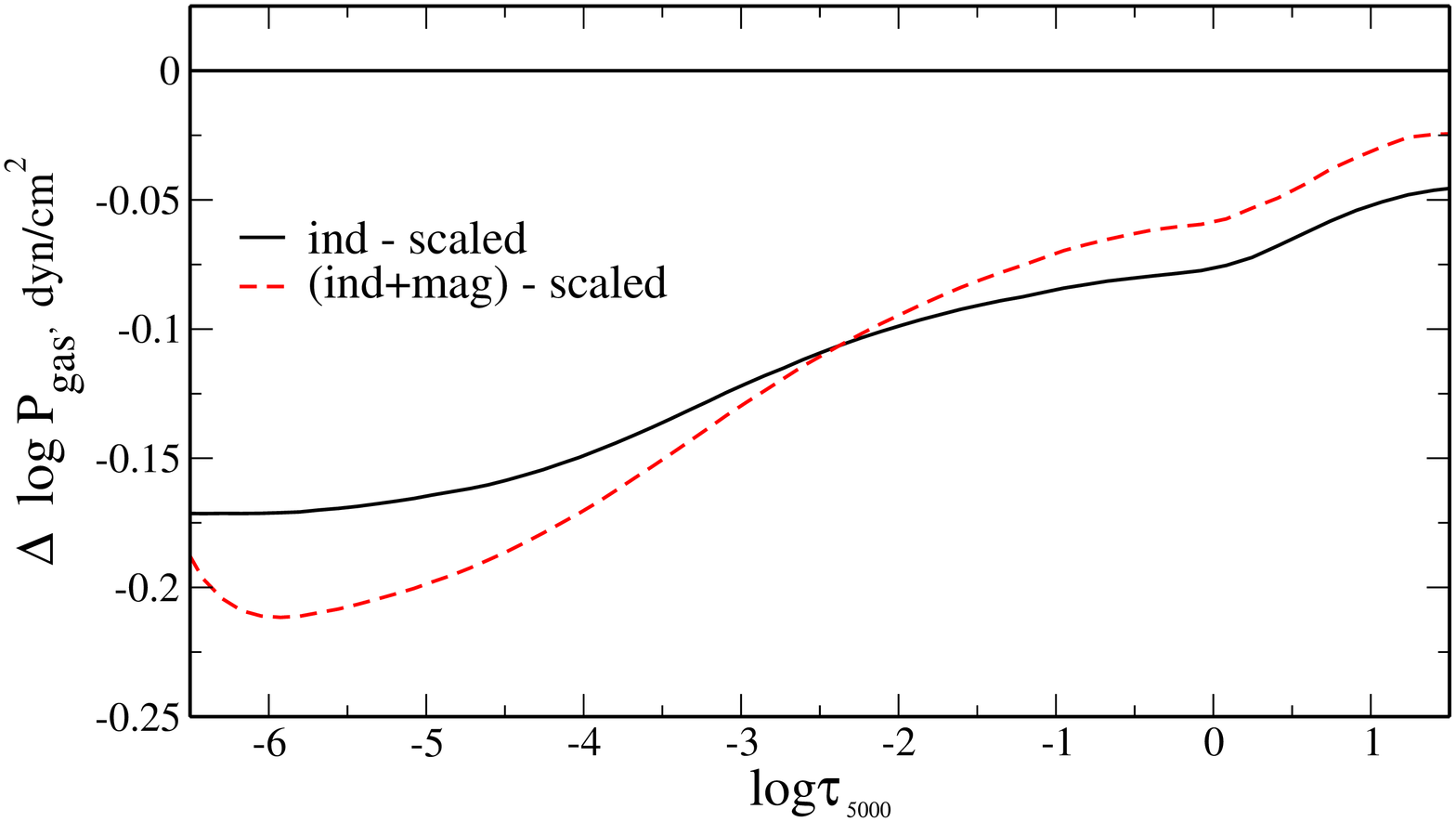}
\caption{Difference in temperature ({\it upper panel}) and gas pressure ({\it bottom panel}) of the models
calculated with individual abundances (solid line) and with individual abundances$+$magnetic field
(dashed line) with respect to the reference model atmosphere computed with scaled-solar abundances
($[M/H]=+1$). For all models $\teff=12\,750$\,K and $\logg=3.8$ is adopted.}
\label{Ftp}
\end{figure}

\begin{figure}[!t]
\resizebox{\hsize}{!}{\includegraphics*{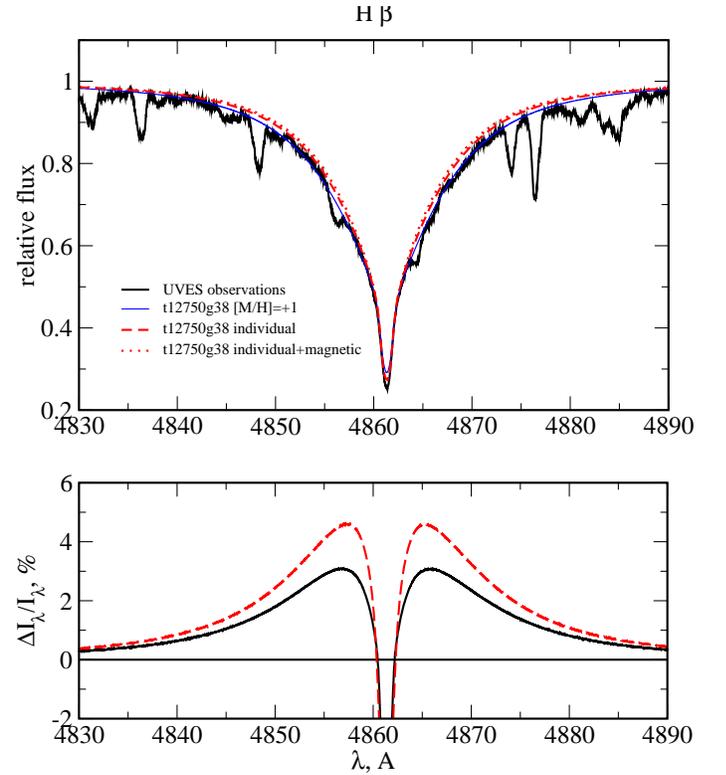}}
\caption{Observed and calculated \hbeta\, line profiles. {\it Upper panel}: thick line~--~UVES
observations (\paper1),
thin line~--~the $\logg=3.8$ model with solar-scaled abundances $[M/H]=+1$,
dashed line~--~the model with individual abundances,
dotted line~--~the model with both individual abundances and the magnetic field.
%dash-dotted line~--~the same as previous, but with $\logg=4.0$.
{\it Bottom panel}: difference (in per sent) between model with only individual abundances (solid line)
and with individual abundance$+$magnetic field (dotted line) relative to the scaled-solar abundance model
($\teff=12\,750$\,K, $\logg=3.8$ for all models).}
\label{Fhbeta}
\end{figure}

\begin{figure}[!t]
\resizebox{\hsize}{!}{\includegraphics*{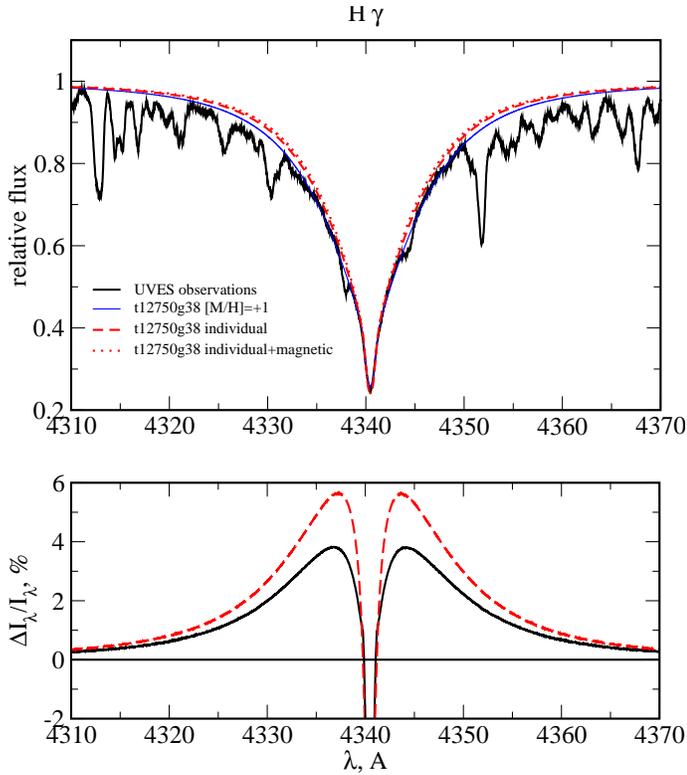}}
\caption{Same as in Fig.~\ref{Fhbeta} but for the \hgamma\, line.}
\label{Fhgamma}
\end{figure}

The effect of the peculiar abundance pattern and magnetic field on the temperature-pressure structure of the
model atmosphere of HD\,137509 is shown in Fig.~\ref{Ftp}. The model with individual abundances exhibits a
decrease of temperature in the surface layers, while the layers close to the photosphere ($\log\tau_{\rm
5000}\approx0$) are heated compared to the reference scaled-solar abundance model ($[M/H]=+1.0$). The
effect of magnetic field is, therefore, considerably different compared to a change in metal abundance. The
inclusion of magnetic field leads to heating of the surface layers due to additional opacity in the Zeeman
components. This occurs because a high line density in the dominant UV part of the stellar spectral
energy distribution makes it possible for the backwarming effect to occur even in surface layers which become
opaque at these wavelengths \citep{zeeman_paper1}.
The temperature distribution of the magnetic model in the surface
layers tends to be close to that of  the scaled-solar model. At the same time, magnetic field
appears to influence the temperature structure more efficiently in deep atmospheric layers compared
to the non-magnetic model with individual abundances.

Both magnetic and non-magnetic models show a decrease of the gas pressure throughout the atmosphere
compared to the reference model. However, the model with the magnetic field shows a steeper depth dependence
of the pressure difference relative to the model where only individual abundances are taken into
account. In our hydrostatic models the decrease of the gas pressure is caused by the respective increase of the radiative pressure due to
enhanced opacity for models with individual abundances and magnetic field.
Note that the model with the magnetic field included exhibits a noticeable increase of the radiative pressure 
in the outer atmospheric layers compared to other models. This behaviour of the radiative pressure and 
corresponding changes
in the overall temperature-pressure structure of the magnetic atmosphere are relevant for modern studies
of the radiatively-driven chemical diffusion in the atmospheres of magnetic stars \citep[e.g.,][]{alecian}.

At this point it is difficult to predict the total effect of the magnetic field on the spectral
characteristics of the star. In two different atmospheric regions (surface and photospheric layers),
the magnetic field influences the model structure in a different manner, so that the  properties of
magnetic atmosphere of HD\,137509 are not equivalent to a usual non-magnetic model with a different
set of fundamental parameters. One can expect that spectral features which are selectively sensitive to
either temperature or pressure may show different behavior. In the following section we examine
the overall effect of the individual abundances and magnetic field on the hydrogen Balmer line profiles
and metallic line spectra.

\subsection{Hydrogen line profiles and metallic line spectra}

The introduction of peculiar abundances taken from \paper1\, (see Table~\ref{Tabn}) to the model
atmosphere calculations for HD\,137509, together with the magnetic field, leads
to noticeable changes in the hydrogen line profiles.
In Figs.~\ref{Fhbeta} and~\ref{Fhgamma}, we compare theoretical profiles for different
model atmospheres with those observed for HD\,137509.
Synthetic profiles were calculated using the {\sc Synthmag}
program \citep{synthmag}. This code incorporates recent improvements
in the treatment of the hydrogen line opacity \citep{barklem} and takes into account magnetic
splitting of hydrogen lines. However, possible modification of the Stark broadening by magnetic field
\citep{mathys} is neglected. The observed profiles
of hydrogen lines are extracted from the UVES spectrum of HD\,137509 described in \paper1.

One can see that in order to retain acceptable agreement between observations and theory,
it is necessary to increase the $\log g$ value from $3.8$ derived in \paper1\ to $4.0$.
Note, that most of the changes in the hydrogen line profiles are due to anomalous
abundances adopted for the star (mainly due to extreme He underabundance). Nevertheless,
the changes in the temperature-pressure structure of the atmosphere
produced by the magnetic field are also important and has to be incorporated to the model in
order to retrieve an accurate estimate of the gravitational acceleration.

\begin{figure}
\includegraphics*[angle=90,width=\hsize,height=5cm]{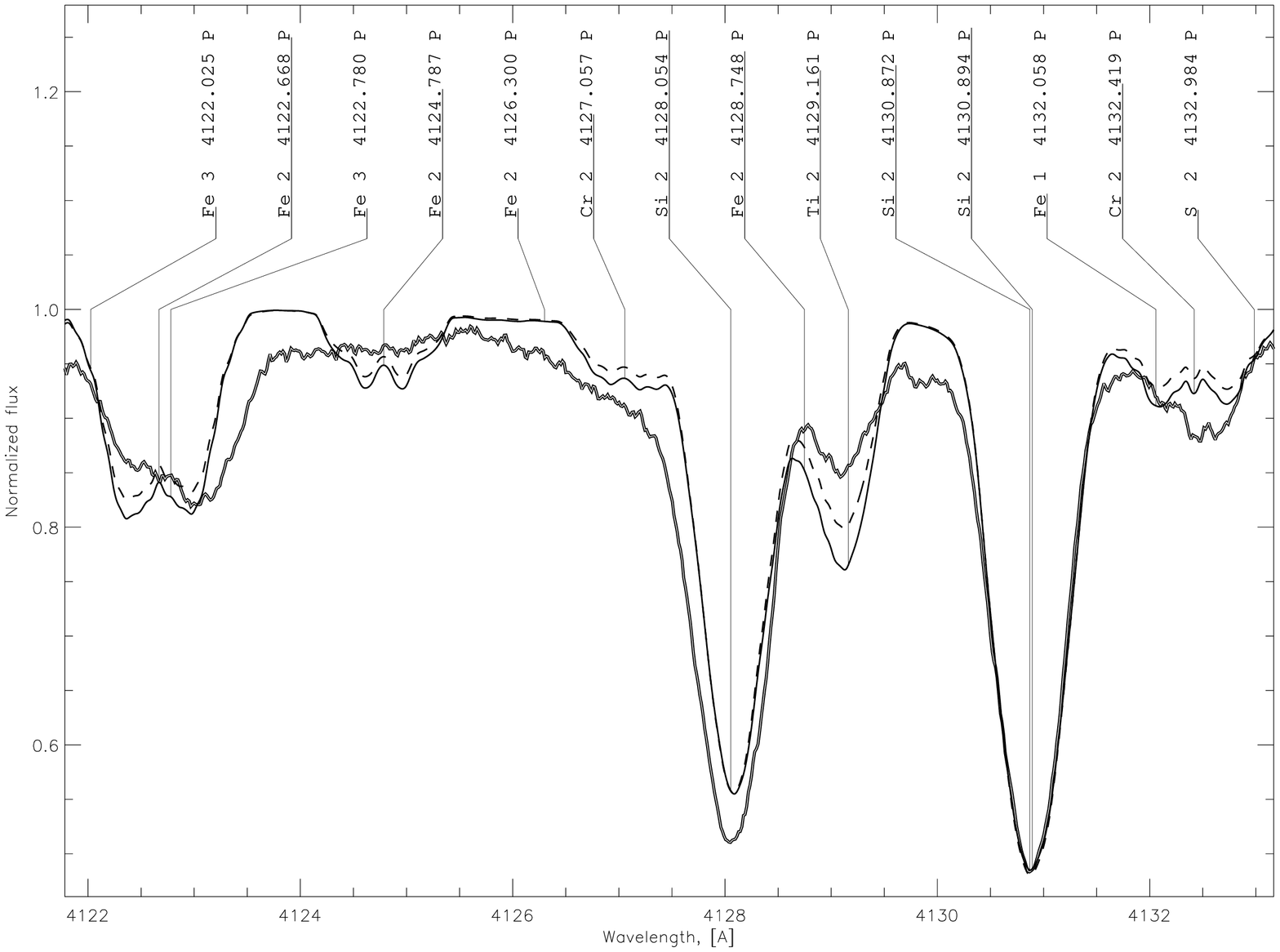}
\includegraphics*[angle=90,width=\hsize,height=5cm]{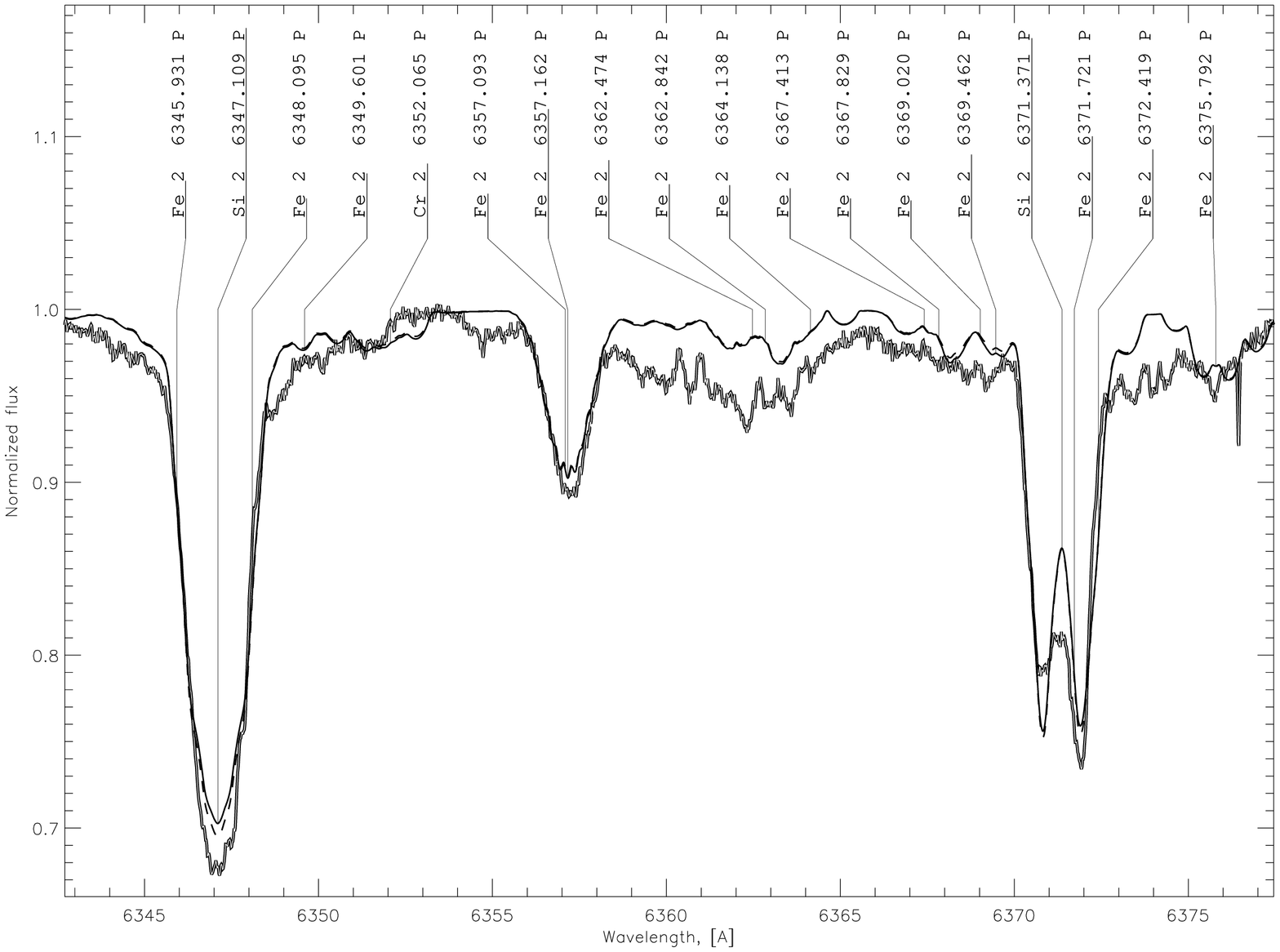}
\caption{Influence of magnetic model structure on the profiles of selected spectral lines for
$\teff=12\,750$\,K, $\logg=4.0$ models.
Double line -- UVES observations, thin line -- spectra calculated using the non-magnetic atmospheric model,
dashed line -- with the magnetic field included.}
\label{Flines}
\end{figure}

The metallic line spectra are not particularly sensitive to the changes in the atmospheric model structure
associated with the inclusion of the magnetic field. Fig.~\ref{Flines} shows the observed and synthetic profiles
of some prominent spectral lines of silicon. All theoretical spectra were calculated using the {\sc Synthmag}
code with the same abundances and the homogeneous surface magnetic field distribution, characterized by
$\langle B \rangle=29$~kG, but using two different model atmospheres:
one model is only with individual abundances and another one includes both anomalous abundances and the magnetic
field. For both models the fundamental parameters are $\teff=12\,750$\,K, $\logg=4.0$.
The calculated spectra were convolved with $v\sin i=20$~\kms\, rotational broadening to allow
comparison with observational data.
The effects of the magnetic field in the model atmosphere distort different lines in different ways. Some of the
spectral features considered here are not sensitive to the magnetic model atmosphere effects at all
(for example, \ion{Si}{ii} 4130.872\,\AA\ and \ion{Si}{ii} 4130.894\,\AA), while other lines show a noticeable
discrepancy between the spectra computed for magnetic and non-magnetic atmospheres (e.g., \ion{Ti}{ii}
4129.161\,\AA).
The average difference between the line profiles obtained from the magnetic and non-magnetic models was found to be
about few per sent for the spectra already broadened by rotation.
Due to a very large magnetic broadening (comparable to the rotational Doppler effect), the effect on
unconvolved spectra is roughly of the same magnitude for all spectral lines considered. These
results are in a good agreement with our previous investigation \citep{zeeman_paper1}. However, the present
results are more accurate because polarized radiative transfer was taken into account for both the model atmosphere
and the spectrum synthesis calculations.

\subsection{Energy distribution}

Generally, for self-consistent modelling of stellar atmospheres,
the following observables must be reproduced simultaneously: hydrogen line profiles, energy distribution
and metallic line spectra. Among these, the flux distribution is especially sensitive to the overall energy balance in the stellar
atmosphere over a wide range of optical depths. So, this observable is extremely useful for determining basic
stellar parameters and is preferable in this respect to broad- and intermediate-band photometric data.
The role of energy distributions is even more important for chemically peculiar stars because most of
the widely used photometric calibrations are based on
observations of normal stars and, hence, may not be applicable to stars with strong magnetic fields and
unusual abundances \citep{indabn}.

\begin{figure*}
\begin{center}
\includegraphics*[angle=-90,width=14cm,clip]{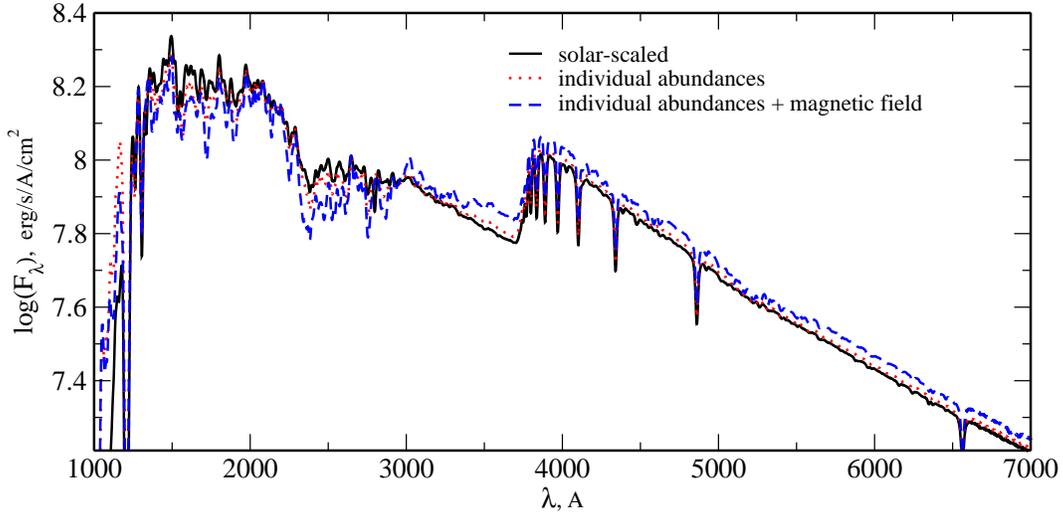}
\end{center}
\caption{Theoretical energy distributions for the models with $\teff=12\,750$\,K, $\logg=3.8$.
Solid line -- scaled-solar abundance model,
dotted line -- the model with individual chemical composition,
dashed line -- the model with individual abundances and the magnetic field.
Energy distributions are convolved with a Gaussian profile with ${\rm FWHM}=15$~\AA.}
\label{Fed3models}
\end{figure*}

Figure~\ref{Fed3models} illustrates the effects of individual abundances and the magnetic field
on the energy distributions for the models of HD\,137509 with $\teff=12\,750$\,K, $\logg=3.8$.
It is evident
that both the model with individual abundances and the model combining the effects of unusual chemical composition
and the strong magnetic field exhibit flux redistribution from the UV to visual wavelength region.
The most significant effect is found for the model
with the magnetic field included, which is in agreement with our previous results.

Due to the increased level of flux in the visual region, the models with magnetic line blanketing are
characterized by an anomalous bolometric correction. We find that the difference in this parameter with respect to the
computations done for solar chemical composition reaches $\Delta {\rm BC}=0.1$~mag for
the model with individual abundance, whereas $\Delta {\rm BC}\approx 0.15$~mag for the model with individual
abundance and magnetic
field. Thus, the anomaly in the bolometric correction is quite substantial and has to be
taken into account in the determination
of absolute luminosity and comparison with evolutionary models of CP stars \citep[e.g.,][]{evol}, yet, even
for such an extreme star as HD\,137509, we cannot confirm the reality of $\Delta {\rm BC}>0.2$~mag
proposed by \citet{lanzbc}.

\begin{figure}
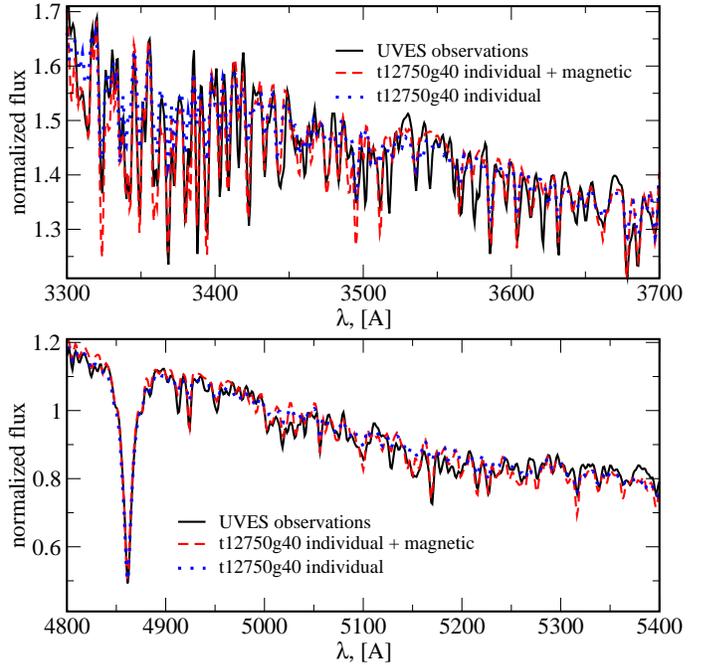

\resizebox{\hsize}{!}{\includegraphics*{figures/fuv.eps}}\\
\resizebox{\hsize}{!}{\includegraphics*{figures/f5000.eps}}
\caption{Comparison between the observed and calculated fluxes of HD\,137509
in the near UV (upper panel) and visual region (lower panel). Gaussian smearing with resolution $R=1500$
is applied to all the spectra to provide a better view.
In the upper plot the fluxes for the non-magnetic model are shifted along the vertical axis
to match the observed spectrum.}
\label{Fed}
\end{figure}

Unfortunately, no suitable observed energy distribution for HD\,137509
which could be used for verifying theoretical models is available
in the literature.
One can possibly use the flux-calibrated spectrum from the UVES Library\footnote{{\tt http://www.eso.org/uvespop}}.
The description of the reduction of these data can be found in \citet{uves}.
Despite the fact that the UVES pipeline provides a user with the spectra calibrated in relative units,
these data can not be used for comparisons with models in a wide
spectral range due to the flux calibration uncertainties (S. Bagnulo, private
communication) that could affect the shape of the energy distribution.
Nevertheless, the relative shape of the UVES spectra within small spectral
regions (300--500 \AA) is reasonably well-determined. Therefore, in Fig.~\ref{Fed} we compare
relative fluxes (normalized at $\lambda=5000$\,\AA) of the magnetic and non-magnetic models of HD\,137509
with the UVES spectrum in two short regions. Note, that both theoretical models presented in the
figure were calculated with the abundances from \paper1. Thus, the difference between the two model
fluxes is entirely due to the magnetic field, which is responsible for producing complex spectral features
clearly seen in the observed spectra. In particular, the Zeeman splitting influences
the strength of the line absorption around $\lambda=5200$\,\AA, which is associated with the
well-known depression in the spectra of CP stars and its photometric characteristic
($\Delta a$ photometry, discussed below).
Note, that for the comparison of fluxes in the UV region (upper panel in Fig.~\ref{Fed}) we had to shift the
energy distribution produced by non-magnetic model along the $y$-axis to match the observed data. This is
a consequence of the substantial difference in the fluxes close to the Balmer jump where the model with the
magnetic field shows a prominent flux excess compared to the non-magnetic model (see Fig.~\ref{Fed3models}).

The only other possibility would be to use the flux calibrated spectra of HD\,137509 obtained by the Far
Ultraviolet Spectroscopic Explorer\footnote{{\tt http://fuse.pha.jhu.edu/}} (FUSE) mission. However, these data
were obtained in a very short wavelength range ($910$--$1185$\,\AA), which makes it impossible to use these spectra
for accurate determination of fundamental stellar parameters and testing theoretical models. For example, the
overall radiative energy emitted inside $910$--$1185$\,\AA\, region is about $1.6$\% of the total flux for the
model with $\teff=12750$\,K and about $2.5$\% for the $\teff=13750$\,K model, which is unimportant for overall
radiative energy balance in the atmosphere of HD\,137509. Moreover, the slope of the energy distribution in this
region remains the same for both models mentioned above. This means that, based on FUSE data, it is impossible  to
distinguish models different by as much as 1000~K. Well-calibrated energy distribution covering a wide spectral
regions is needed.

\subsection{Photometric colors}

As a next step, we have calculated a grid of model atmospheres with different effective
temperatures ($\teff=12\,250$\,K--$13\,750$\,K, $\Delta T=250$\,K)
and gravities ($\logg=3.8$--$4.2$, $\Delta \logg=0.2$) to assess the ability of our models to reproduce
the observed photometric properties of HD\,137509. To investigate the role of magnetic field,
we have included in the grid both magnetic and non-magnetic models with individual abundances.
We also considered the reference scaled-solar abundances
model from \paper1.

The theoretical colors were calculated following the procedure outlined in \citet{zeeman_paper1}. We use
modified computer codes by \citet{kurucz13}, which take into account transmission
curves of individual photometric filters, mirror reflectivity and
a photomultiplier response function. The synthetic $\Delta a$ values were computed
with respect to the theoretical normality line, $a_0$, determined in \citet{indabn}.
The reddening, corresponding to $E(B-V)=0.06$, was taken into account for all color indices,
except $X$, $Y$ and $Z$ which are reddening-free \citep{xyz}. This value of the color excess was found
from the intrinsic $[U-B]$ color and reddening-free Geneva $X$ and $Y$ parameters \citep{cramer}. This estimate
agrees with the $E(B-V)$ range of 0.04--0.08 that one can infer from the dust maps of \citet{dust} and the revised
Hipparcos parallax of HD\,137509, $\pi=5.12\pm0.38$ mas \citep{hippar}.
The observed Str\"omgren and Geneva photometric parameters of
HD\,137509 were taken from the catalogues of \citet{stromgren} and \citet{geneva}, respectively. The observed
value of the $\Delta a$ parameter was adopted from \citet{paunzen}. These photometric characteristics of
HD\,137509 are summarized in the first row of Table~\ref{Tcolors}, which also shows the results of the synthetic color calculations.
We present results only for those models which fit the \hbeta\, and \hgamma\, line profiles reasonably well.
We have also looked into possibility of using infra-red photometric measurements (e.g.,
2MASS\footnote{\tt http://www.ipac.caltech.edu/2mass/} or observations by \citet{irphot}) to constrain the
model parameters. However, the difference between all models fluxes of HD\,137509 considered below
is less than the observational uncertainty in this spectral range.

\begin{table*}
\caption{Observed and reddening-corrected calculated photometric parameters of HD\,137509.
Theoretical values were calculated
for the models with (``mag'') and without (``non'') magnetic field, using individual abundances. Predictions
for the scaled-solar abundance non-magnetic model of \paper1\ are given for comparison.
The last row reports predictions for the magnetic and non-magnetic models with revised abundances
determined in the present paper (see text)}.
\centering
\begin{tabular}{lccccccccccccc}
\hline\hline
                        &          $b-y$    &        $m_{\rm 1}$ &         $c_{\rm 1}$ &        $\Delta a$ &          $U$--$B$ &      $V$--$B$     &       $G$--$B$    & $B_{\rm 1}$--$B$  & $B_{\rm 2}$--$B$  & $V_{\rm 1}$--$B$  &            $X$    &            $Y$    &           $Z$     \\
\hline
\textbf{observations}   & $\mathbf{-0.095}$ & $\mathbf{ 0.183}$  & $\mathbf{ 0.411}$   & $\mathbf{ 0.066}$ & $\mathbf{ 0.761}$ & $\mathbf{ 1.136}$ & $\mathbf{ 2.294}$ & $\mathbf{ 0.811}$ & $\mathbf{ 1.566}$ & $\mathbf{ 1.836}$ & $\mathbf{ 0.762}$ & $\mathbf{ 0.076}$ & $\mathbf{-0.067}$\\
\hline
t12250g38, mag          &         $-0.054$  &         $ 0.142$   &         $ 0.640$    &         $ 0.057$  &         $ 0.995$  &         $ 1.088$  &         $ 2.265$  &         $ 0.838$  &         $ 1.574$  &         $ 1.801$  &         $ 1.026$  &         $ 0.122$  &         $-0.045$\\
t12250g38, non          &         $-0.040$  &         $ 0.134$   &         $ 0.653$    &         $ 0.033$  &         $ 1.010$  &         $ 1.069$  &         $ 2.261$  &         $ 0.846$  &         $ 1.564$  &         $ 1.777$  &         $ 1.046$  &         $ 0.073$  &         $-0.028$\\
\hline
t12750g40, mag          &         $-0.057$  &         $ 0.141$   &         $ 0.572$    &         $ 0.054$  &         $ 0.923$  &         $ 1.094$  &         $ 2.277$  &         $ 0.838$  &         $ 1.574$  &         $ 1.806$  &         $ 0.937$  &         $ 0.088$  &         $-0.044$\\
t12750g40, non          &         $-0.044$  &         $ 0.135$   &         $ 0.586$    &         $ 0.031$  &         $ 0.941$  &         $ 1.075$  &         $ 2.271$  &         $ 0.846$  &         $ 1.565$  &         $ 1.783$  &         $ 0.957$  &         $ 0.045$  &         $-0.029$\\
\hline
t13250g40, mag          &         $-0.059$  &         $ 0.135$   &         $ 0.516$    &         $ 0.051$  &         $ 0.860$  &         $ 1.105$  &         $ 2.293$  &         $ 0.834$  &         $ 1.578$  &         $ 1.815$  &         $ 0.863$  &         $ 0.077$  &         $-0.041$\\
t13250g40, non          &         $-0.046$  &         $ 0.130$   &         $ 0.529$    &         $ 0.030$  &         $ 0.876$  &         $ 1.087$  &         $ 2.285$  &         $ 0.841$  &         $ 1.570$  &         $ 1.794$  &         $ 0.880$  &         $ 0.040$  &         $-0.028$\\
\hline
t13750g42, mag          &         $-0.062$  &         $ 0.138$   &         $ 0.455$    &         $ 0.046$  &         $ 0.798$  &         $ 1.106$  &         $ 2.299$  &         $ 0.839$  &         $ 1.574$  &         $ 1.815$  &         $ 0.780$  &         $ 0.029$  &         $-0.040$\\
t13750g42, non          &         $-0.051$  &         $ 0.138$   &         $ 0.468$    &         $ 0.030$  &         $ 0.815$  &         $ 1.081$  &         $ 2.282$  &         $ 0.850$  &         $ 1.561$  &         $ 1.788$  &         $ 0.790$  &         $-0.023$  &         $-0.030$\\
\hline
t12750g38, mag          &         $-0.056$  &         $ 0.135$   &         $ 0.573$    &         $ 0.054$  &         $ 0.919$  &         $ 1.101$  &         $ 2.283$  &         $ 0.833$  &         $ 1.579$  &         $ 1.812$  &         $ 0.936$  &         $ 0.107$  &         $-0.043$\\
t12750g38, non          &         $-0.043$  &         $ 0.129$   &         $ 0.588$    &         $ 0.032$  &         $ 0.936$  &         $ 1.082$  &         $ 2.277$  &         $ 0.841$  &         $ 1.570$  &         $ 1.790$  &         $ 0.957$  &         $ 0.066$  &         $-0.028$\\
t12750g38, scaled       &         $-0.026$  &         $ 0.116$   &         $ 0.581$    &         $ 0.014$  &         $ 0.932$  &         $ 1.061$  &         $ 2.264$  &         $ 0.847$  &         $ 1.564$  &         $ 1.766$  &         $ 0.942$  &         $ 0.032$  &         $-0.016$\\
%\hline
%$^\ast$t13750g42, mag   &         $-0.082$  &         $ 0.157$   &         $ 0.450$    &         $ 0.061$  &         $ 0.791$  &         $ 1.125$  &         $ 2.311$  &         $ 0.836$  &         $ 1.577$  &         $ 1.836$  &         $ 0.779$  &         $ 0.045$  &         $-0.052$\\
%$^\ast$t13750g42, non   &         $-0.065$  &         $ 0.148$   &         $ 0.467$    &         $ 0.039$  &         $ 0.813$  &         $ 1.100$  &         $ 2.296$  &         $ 0.844$  &         $ 1.568$  &         $ 1.808$  &         $ 0.799$  &         $ 0.005$  &         $-0.038$\\
\hline
$^\ast$t13750g42, mag   &         $-0.081$  &         $ 0.153$   &         $ 0.453$    &         $ 0.062$  &         $ 0.800$  &         $ 1.136$  &         $ 2.321$  &         $ 0.829$  &         $ 1.585$  &         $ 1.847$  &         $ 0.799$  &         $ 0.081$  &         $-0.051$\\
$^\ast$t13750g42, non   &         $-0.064$  &         $ 0.144$   &         $ 0.470$    &         $ 0.037$  &         $ 0.820$  &         $ 1.111$  &         $ 2.307$  &         $ 0.837$  &         $ 1.576$  &         $ 1.819$  &         $ 0.818$  &         $ 0.040$  &         $-0.036$\\
\hline
\end{tabular}
\label{Tcolors}
\end{table*}

There are several things to
note in Table~\ref{Tcolors}. First, it is evident that the change in photometric parameters associated with the
introduction of the magnetic field and individual abundances
in the model atmosphere calculations amounts to a few hundreds of a magnitude compared with the scaled-solar
abundances model and
that, generally, magnetic models allow us to improve the fit to the observed values of all photometric indices.
The only exceptions are the parameter $Y$ for the model with $\teff=12\,250$\,K, $\logg=3.8$ (its value for
non-magnetic model is closer to the observed one compared to the magnetic model) and the index $B_{\rm 2}-B$
(which tends to increase if the magnetic field is introduced to calculations, while decreasing is needed to fit the
observations).  Although the overall changes in the color indices are not very large even for such a strong
magnetic field, we conclude that taking into account Zeeman splitting and magnetic intensification of spectral lines
clearly is a necessary ingredient for modelling photometric observables of magnetic stars. Of course, the dependence of
photometric colors on the magnetic field is sensitive to the chemical composition of the star.
The increased metal content of the stellar atmosphere has a large impact on the photometric colors and thus
HD\,137509 represents one of the extreme cases due to the strong overabundance of  iron-peak elements in its
atmosphere.

To distinguish between the effects of magnetic field and individual abundance we also compare synthetic colors for the models calculated with the
same fundamental parameters ($\teff=12750$\,K, $\logg=3.8$) but different assumptions about abundances and magnetic field. Comparing the models
with scaled-solar abundances with individual abundances model and the latter model with the magnetic model
atmosphere$+$individual abundances, one can infer that the effect of the very strong magnetic field on photometric
observables is generally
comparable in magnitude with the effect of  individual abundances alone. Interestingly, for some of the color
indices the difference between the models  calculated with and without magnetic field is larger than the same
difference between the non-magnetic model with individual abundances and the reference  scaled-solar model. This is
the case for $c_{\rm 1}$, $\Delta a$, $U-B$, $X$, $Y$ and $Z$. One should not forget that we are dealing
with an effect depending on many parameters, i.e. the overall picture will depend upon the effective temperature
and the abundance pattern, and thus could differ a lot from one star to another even if the field strength is the
same. Limiting our conclusions to HD\,137509, we find that the effect of the magnetic field is of no less importance
than the effect of individual abundances for the majority of the photometric colors considered in the present study.

Finally, we have also investigated the importance of accurate photoionization cross-sections for individual states of
\ion{Si}{i} and \ion{Si}{ii} ions taken from the TOPBase database \citep{topbase}. The importance of accurate
photoionization cross-sections with full resonance structure of \ion{Si}{ii} was noted by \citet{lanz}. They
found that accounting for wide TOPBase resonances in the spectrum synthesis calculations leads to a
good fit to the observed UV spectra of the Bp star HD\,34452 ($\teff=13\,650$\,K,
$\logg=4.0$), which has an enhanced Si abundance. This could be an important effect for HD\,137509 because this
star is also characterized by an excess of Si in its atmosphere. Thus, we have implemented available \ion{Si}{i}
and \ion{Si}{ii} cross-section data from the TOPBase  web page\footnote{{\tt
http://vizier.u-strasbg.fr/topbase/topbase.html}} in the \llm\ atmospheric calculations.  In this modelling we
did not use any additional techniques, such as resonance-averaging of photoionization cross-sections data or
combining energy levels to superlevels which are usually suggested to reduce computational expenses.
All the energy levels for both ions
were explicitly included in the opacity calculations. We found that even for the hottest model considered here
the effect of detailed photoionization cross-sections is negligible for the majority of color indices in comparison
with the bound-bound treatment of resonance structure via autoionizing lines. The
strongest deviation between the model with the TOPBase data and the model with  \ion{Si}{i} and \ion{Si}{ii}
continuous opacities taken from the {\sc Atlas12} code \citep{kuruczA12} is found for $c_{\rm 1}$
($-0.004$\,mag), $U-B$ ($-0.007$\,mag), $X$ ($-0.01$\,mag) and $Y$ ($-0.01$\,mag). We found that the main reason for
such a small influence of the detailed photoionization data is that some of the
strong autoionizing lines of \ion{Si}{i} and \ion{Si}{ii} are already included in the VALD database and hence in
the standard \llm\ input line list.  This means that the photoionization cross-sections from {\sc Atlas12} together
with the autoionizing lines from VALD provide a reasonably good approximation to the opacity calculated from
the more accurate TOPBase data, at least for temperatures around $12\,750$\,K. It is worth to note that here we
are mainly interested in the color indices, however, the accurate photoionization cross-sections
are essential when it comes to the comparison of theoretical calculations with high-resolution UV spectra
of Si-rich stars.

\subsection{An improved model for HD\,137509}

The behavior of $b-y$, $c_{\rm 1}$ and $X$ indices suggests that the effective temperature of HD\,137509
could be higher than the initially derived value of $12\,750$\,K which was used in \paper1. To improve the model
parameters in a consistent way we redetermined abundances of chemical elements using model atmosphere with
$\teff=13750$\,K, $\logg=4.2$ calculated with magnetic field and individual abundances from the second column of
Table~\ref{Tabn}.  The resulting new abundances, determined using the same methodology and spectral regions as in
\paper1,  are listed in the third column of Table~\ref{Tabn}. 
The new chemical abundances are systematically 
higher by $0.2$--$0.4$~dex compared to the values reported in \paper1. 
This is mostly a result of using the 
model atmosphere with $1000$\,K higher effective temperature.
The atomic lines used for abundances analysis become systematically shallower with the increase of 
atmospheric temperature thus resulting in positive abundance corrections.
Using these new abundances, we made a next iteration by re-calculating model atmosphere
to ensure consistency of the model temperature-pressure structure and adopted abundances.  This final model
allowed us to fit the hydrogen line profiles as good as previous models. It also improves, with a few exceptions,
the agreement between
observed and computed photometric parameters as demonstrated by the two last rows of Table~\ref{Tcolors}  (models
marked with asterisks).
Thus, we can conclude that the effective temperature of HD\,137509 should be around $13750$\,K.

We would like to note that exact fitting of the photometric indices is
hardly feasible for any model  due to additional unknown parameters involved in the modeling
process. One of the effects which we have ignored is  inhomogeneous surface
distribution of chemical elements in HD\,137509. \citet{mathys_lanz} discussed rotational
modulation of the equivalent widths of  \ion{Cr}{ii}, \ion{Si}{ii} and \ion{Fe}{ii} lines,
and the light variability in the $U$, $B$ and $V$ filters with the maximum amplitude of about
0.05~mag for the $U$ filter. These phenomena are usually attributed to the presence of
abundance spots on the stellar surface. Therefore, to fully characterize atmospheric
properties of HD\,137509, one would need to obtain time-resolved observations and reconstruct
surface distribution of magnetic field and chemical abundances with the Doppler imaging
technique \citep[e.g.,][]{mdi}. Furthermore, the vertical stratification of chemical elements
\citep[e.g.,][]{strat} could also take place in the atmosphere of HD\,137509. A refined
analysis of these chemical inhomogeneity effects is outside the scope of our study.

\section{Summary and conclusions}
\label{concl}

In the present paper we have constructed advanced theoretical stellar model atmospheres
incorporating accurate treatment of the individual abundances pattern, Zeeman splitting, polarized radiative transfer
and compared the results with the observations of extreme magnetic CP star HD\,137509. With the mean surface field of
$\langle B \rangle \approx 29$\,kG, this object has the second largest magnetic field among CP stars.
Strong overabundance of iron-peak elements and extreme underabundance of helium in the atmosphere
of this star opens a possibility to investigate the importance of taking individual abundances into account when
constructing model atmospheres of magnetic chemically peculiar stars. Theoretical model atmosphere calculations
were compared with the hydrogen line profiles and metallic spectrum of HD\,137509. The Str\"omgren and Geneva
photometric parameters were also investigated.
The main conclusions of our study can be summarized as follows:
\begin{itemize}
\item
We found that the effect of individual abundances dominates the change of Balmer
\hbeta\, and \hgamma\, line profiles compared to the model with solar-scaled abundances.
This implies that once the abundances of the star have been determined using an approximate
model, it is necessary to recalculate the model atmosphere in order to ensure the consistency between
abundance pattern and the model structure. The magnetic field has less influence on the hydrogen
lines, however it should be taken into account for stars with very strong magnetic fields.
\item
Modification of the atmospheric temperature-pressure structure due to presence of the magnetic field and peculiar abundances
generally has little impact of the metal line profiles comparing to that of non-magnetic one.
\item
For HD\,137509, the effect of magnetic field on photometric colors
is very important for some photometric parameters. Occasionally the combined impact of the magnetic field
and the realistic chemistry
is more important
than the effect of using only individual abundances. Generally, magnetic model atmospheres
allowed us to obtain a better agreement between
almost all observed and theoretical color indices. Thus, we can conclude that
magnetic field should be taken into account in the analysis of stars with strong magnetic fields.
\item
We showed that the analysis of the spectra of such extreme Bp stars with strong magnetic fields and unusual
chemistry as HD\,137509 requires a self-consistent approach. Once the abundances of the most important elements
are derived using an approximate model atmosphere,
it is necessary to recompute the model with new abundances trying to fit various observables such as
hydrogen line profiles, photometric colors, and energy distribution (if available). For HD\,137509 we found that the
simultaneous fit to the hydrogen line profiles
and photometrical indices employing the model with both magnetic field and individual abundances included
requires as much as $1000$\,K
correction to the effective temperature and $0.4$~dex correction to the surface gravity
compared to the results obtained using simple scaled-solar models.
\item
We expect that the overall energy distribution of the star is strongly modified by magnetic
line blanketing. However, the lack of accurate energy distribution covering meaningful wavelength region
in the spectrum of HD\,137509 precludes us from reaching robust quantitative conclusions. We emphasize that availability
of the flux distributions for this and other magnetic CP stars is of great importance for the stellar parameter
determination and for verification of the new generation model atmospheres of magnetic CP stars.
\end{itemize}

\begin{acknowledgements}
This work was supported by
FWF Lisa Meitner grant Nr. M998-N16 to DS,
Postdoctoral Fellowship at UWO funded by a Natural Science Engineering Council
of Canada Discovery Grant to SK and
Austrian Science Fund (FWF) P17890 to OK.
We also acknowledge the use of data
from the UVES Paranal Observatory Project (ESO DDT Program ID 266.D-5655),
FUSE electronic database operated for NASA by the Johns Hopkins University
under NASA contract NAS5-32985
and electronic databases (VALD, SIMBAD, NASA's ADS).
\end{acknowledgements}

\end{document}